\documentclass[%
 aps, prl, reprint,
 superscriptaddress, 
 showpacs, nopreprintnumbers,
 longbibliography,
 amsmath, amssymb, floatfix
]{revtex4-2}

\usepackage[T1]{fontenc}
\usepackage{graphicx}% Include figure files
\usepackage{dcolumn}% Align table columns on decimal point
\usepackage{bm}% bold math
\usepackage{hyperref}% add hypertext capabilities
\usepackage{physics}% Defines \ket, \bra, \dyad etc.
\usepackage[usenames,dvipsnames]{xcolor}
\usepackage{tabularx}
\usepackage{amsmath}
\usepackage{booktabs}

\newcommand{\mytitle}{Realizing multiband states with ultracold dipolar quantum simulators}

\begin{document}

\title{\mytitle}

\author{Yuliya Bilinskaya}
  \email{yubi9221@student.su.se}
\affiliation{Department of Physics, Stockholm University, AlbaNova University Center, 106 91 Stockholm, Sweden}

 \author{Michael Hughes}
  %\email{}
  \affiliation{Clarendon Laboratory, University of Oxford, Parks Rd, Oxford OX1 3PU, United Kingdom.}
  
\author{Paolo Molignini}
  \email{paolo.molignini@fysik.su.se}
\affiliation{Department of Physics, Stockholm University, AlbaNova University Center, 106 91 Stockholm, Sweden}

\date{\today}

\begin{abstract}
% Simulation of dipolar interaction with ultracold molecules has become a significant recent development towards realising the ultimate goal of simulating quantum systems beyond contact interaction. A range of promising phases arising due to the dipolar interaction have already been theoretically postulated \cite{Kora:2020, Staudinger:2023, Morera:2023, Bland:2022}. As we are entering a new era of long-range simulations a natural question arises: are all phases realisable by a dipolar quantum simulator using ultracold atoms or molecules? In order to answer this question we have looked at a wide range of parameter regimes of the single-band Dipolar Bose-Hubbard Model – an extension of one of the most commonly used theoretical frameworks in condensed matter \cite{Fisher:1989,Jaksch:1998,Jaksch:2005}. Our findings indicate that the interplay between shallow optical lattice, strong dipolar interaction and higher particle fillings either reduces the fidelity drastically or generates a state with fundamentally different physical properties compared to the target state. Further by systematically checking the parameter regimes we also find regimes where the fidelity is exceptionally high and the target state is experimentally accessible.

% PREVIOUS ABSTRACT
The manipulation of dipolar interactions within ultracold molecular ensembles represents a pivotal advancement in experimental physics, aiming at the emulation of quantum phenomena unattainable through mere contact interactions.
%In this study, we elucidate the capacity of such experimental setups to engender phases that transcend the conventional lowest-band lattice framework. 
Our study uncovers regimes of multiband occupation which allow to probe more realistic, complex long-range interacting lattice models with ultracold dipolar simulators.
By mapping out experimentally relevant ranges of potential depths, interaction strengths, particle fillings, and geometric configurations, we calculate the agreement between the state prepared in the quantum simulator and a target lattice state. 
We do so by separately calculating numerically exact many-body wave functions in the continuum and single- or multiband lattice representations, and building their many-body state overlaps.
Our findings reveal that for shallow lattices and stronger interactions above half filling, multiband population increases, resulting in fundamentally different ground states than the ones observed in simple lowest-band descriptions, e.g. striped vs checkerboard states.
%Our findings indicate that, for shallow lattices, stronger interactions, and in particular above half filling, multiband population increases, leading to states with different physical properties.
%Furthermore, we show that the interplay between commensurability and interactions can lead to quasidegeneracies, rendering a faithful ground state preparation even more challenging.
A wide range of probed parameter regimes in its turn provides a systematic and quantitative blueprint for realizing multiband states with two-dimensional quantum simulators employing ultracold dipolar molecules.
\end{abstract}
\maketitle

\textit{Introduction} --- 
Ultracold atomic and molecular systems have emerged as fundamental tools for exploring quantum physics and simulating complex many-body phenomena~\cite{Zwerger:2003,Bloch:2008, Esslinger:2010,Bloch:2012,Ritsch:2013,Langen:2015,Gross:2017,Blackmore:2018,Cooper:2019,Schaefer:2020,Mivehvar:2021}. 
The ability to control and manipulate magnetic atoms~\cite{Ferrier-Barbut:2016, Baier:2016, Norcia:2021-NatPhys, Chomaz:2023} and dipolar molecules~\cite{Lahaye:2009, Molony:2014, Reichsoellner:2017, Gadway:2016, Moses:2017, Stevenson:2023, Gregory:2024} has dramatically broadened the spectrum of quantum systems we can study, particularly those governed by long-range interactions. 
This has led to significant breakthroughs, including the creation of molecular dipolar Bose-Einstein condensates~\cite{Bigagli:2024}, supersolid phases~\cite{Boettcher:2019, Tanzi:2019, Tanzi:2019-2, Chomaz:2019, Guo:2019, Natale:2019, Tanzi:2021, Norcia:2021, Sohmen:2021, Schmidt:2022,  Sanchez-Baena:2023, Recati:2023, Su:2023}, and quantum magnets~\cite{Li:2023,Bao:2023,Christakis:2023}. 
Additionally, when integrated with optical lattices, these systems allow for the simulation of long-range Hamiltonians akin to those in condensed matter and high-energy physics, unveiling unique behaviors not seen in traditional systems with short-range interactions. 
This progress in experimental capabilities is matched by theoretical advancements, enhancing our understanding of quantum systems' properties and dynamics and uncovering novel phenomena such as new quantum and spin liquid phases, altered phase transition behaviors, and modified dynamics and entanglement spread~\cite{Chen:2019, Machado:2020, Defenu:2023, Morera:2023}.

A particularly exciting development in this field is employing interactions to more readily construct complex extensions of prototypical lattice models such as the Bose-Hubbard model (BHM)~\cite{Fisher:1989,Jaksch:1998,Jaksch:2005}, a cornerstone in the study of strongly correlated physics~\cite{Gersch:1963,Das:1999,Yurkevich:2001,Fazio:2001,Greiner:2002,Bruder:2005,Bakr:2010,Takafumi:2017,Lin:2019,Lin:2021,Oosten:2003,Larson:2009,Dutta:2011,Mering:2011,Luehmann:2012,Martikainen:2012,Lacki:2013,Xu:2016,Li:2023-PRA,Hughes:2022,Hughes:2023}.
While research has largely focused on the BHM's adaptation to long-range interactions~\cite{Xie:2004,Dutta:2015,Biedron:2018,Lagoin:2022,Tamura:2022} and the new quantum phases that emerge from them~\cite{DallaTorre:2006,Giamarchi:2008,Pollet:2010,Capogrosso:2010,Zapf:2014,Marciniak:2023}, the potential of dipolar simulators to explore multiband models has received less attention. 
In our study, we precisely identify the conditions --- spanning potential depths, interaction strengths, particle fillings, and geometric configurations --- under which these multiband models can be realized. 
We provide a detailed comparison of the physics observed in experimental setups to that in the corresponding lattice models, assessing their congruence through ground-state energies, density distributions, and many-body wave function fidelities. 
Our findings reveal that for shallow lattices and stronger interactions above half filling, higher bands become rapidly populated, resulting in fundamentally different ground states than the ones observed in simple lowest-band descriptions, e.g. striped vs checkerboard states.
This discrepancy is especially pronounced in incommensurate geometries with nearly degenerate ground states. 
Our work highlights the potential to simulate more intricate multiband lattice models with ultracold dipolar systems, offering insights for future experimental designs and a roadmap for evaluating dipolar quantum simulators' accuracy.

%%%%%%%%%%%%%%%%%%%%%%%%%%%%%
\begin{figure*}
    \centering
     \includegraphics[width=\textwidth]{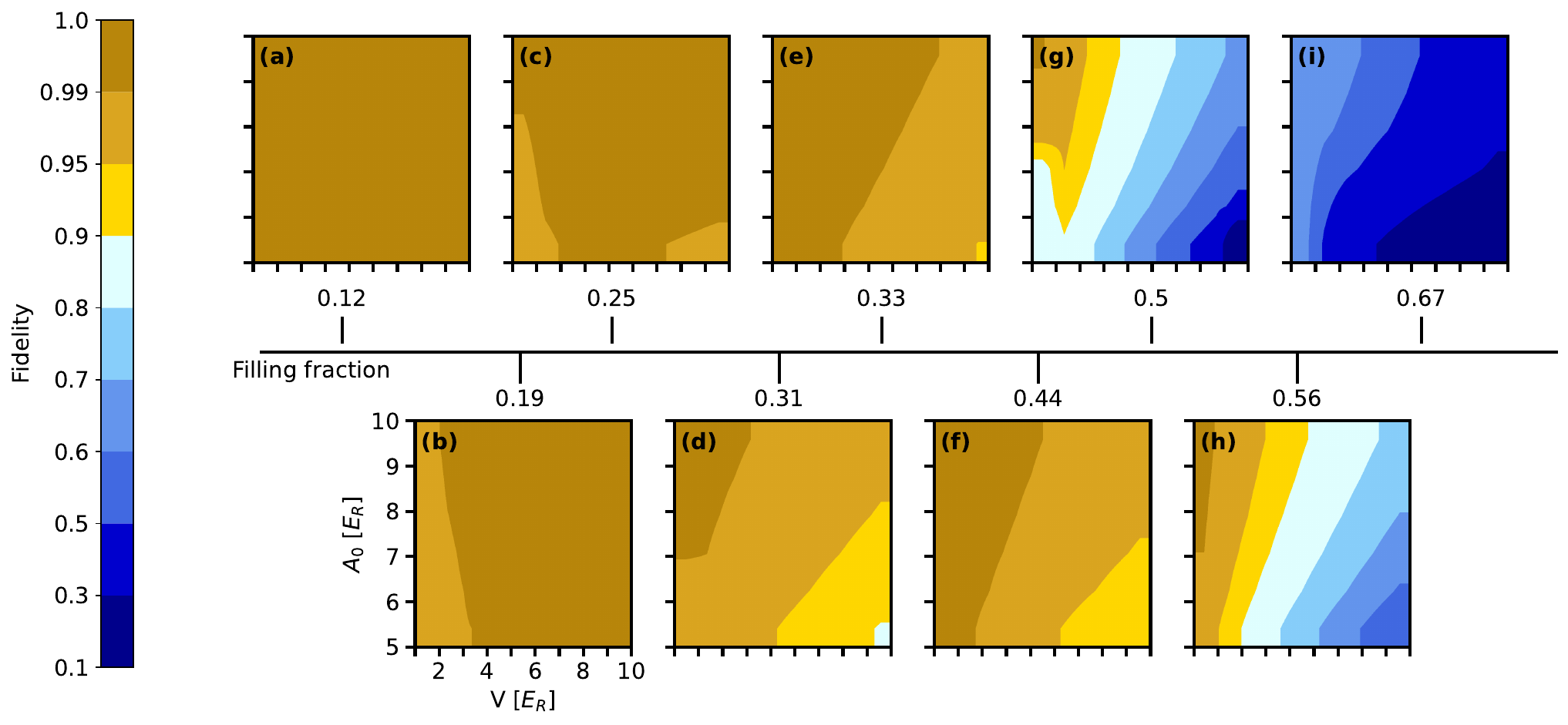}
    \caption{\textbf{Fidelity of 2D dipolar quantum simulators and 1BDBH model in different parameter regimes.} (a)-(i) Many-body fidelity as a measure for the transition of a single- to multi-band dipolar Bose-Hubbard model quantum simulator, for varying 
     lattice depth $A_0$, interaction strength $V$ and filling fraction $\nu$. 
     The labels of (b) apply to all panels. }
    \label{fig:phase-diagrams}
\end{figure*}
%%%%%%%%%%%%%%%%%%%%%%%%%%%%%

%%%%%%%%%%%%%%%%%%%%%%%%%%%%%%%%%%%%%%%%%%%%%%%%%%%%%%%%%%%%%%%%%%%%%%%%%%%%%%%%%%%%%%%%%%%%%%%%%%%%%%%%
\textit{Physical scenario} ---
We consider repulsive-interacting dipolar bosons of mass $m$ in a 2D optical lattice
\begin{equation}
A(\mathbf{x}) = \frac{A_0}{E_R} \left[ \sin^2 \left( \frac{\pi x}{L_0} + \phi_x \right) + \sin^2  \left( \frac{\pi y}{L_0} + \phi_y \right) \right]
\end{equation}
where $L_0$ is the distance between two neighboring minima (unit of length) and $E_R$ is the recoil energy of the optical lattice (unit of energy).
The optical lattice has hard-wall boundaries that restrict the number of minima to exactly $S_x$ in $x$ direction and $S_y$ in $y$ direction ($S_i \in \mathbb{N}$), thereby recreating a lattice with $S_x \times S_y$ sites~\footnote{For an even number of sites in the $j$ direction, we need to set $\phi_j = \pi/2$}.
For the dipole-dipole interactions (DDI), we assume a strong transverse harmonic confinement which suppresses the divergence~\cite{Sinha:2007,Fischer:2015,Chatterjee:2018,Chatterjee:2019,Chatterjee:2020} and yields the regularized form
\begin{equation}
\mathcal{U}_V(\mathbf{x}-\mathbf{x}') = \frac{V L_0^3}{E_R (|\mathbf{x}-\mathbf{x}'|^3 + \alpha)},
\end{equation}
where $\alpha = 0.05$ and $V$ denotes the bare DDI strength between two bosons at unit distance.
To limit multiple occupation of each site to $<1\%$, we add very strong contact repulsions shaped as a narrow Gaussian $\mathcal{U}_G(\mathbf{x},\mathbf{x}')$~\cite{supmat}.
The full system is thus governed by the Hamiltonian
\begin{align}
&\mathcal{H}_{\text{cont}} = 
\int \mathrm{d}\mathbf{x} ~ \hat{\Psi}^{\dagger}(\mathbf{x}) \left[ -\frac{\hbar^2}{2m}\boldsymbol{\nabla}^2 + A(\mathbf{x}) \right] \hat{\Psi}(\mathbf{x}) \nonumber \\
&+ \frac{1}{2}\int \mathrm{d}\mathbf{x} \int \mathrm{d}\mathbf{x}' \: \hat{\Psi}^{\dagger}(\mathbf{x}) \hat{\Psi}^{\dagger}(\mathbf{x}') 
W(\mathbf{x}, \mathbf{x}') \hat{\Psi}(\mathbf{x}') \hat{\Psi}(\mathbf{x}).
\label{eq_continuumH_maintext}
\end{align}
Here $\hat{\Psi}^{(\dagger)}(\mathbf{x})$ creates (annihilates) a boson at position $\mathbf{x}$ and $W(\mathbf{x}, \mathbf{x}') = \mathcal{U}_V(\mathbf{x},\mathbf{x}') + \mathcal{U}_G(\mathbf{x},\mathbf{x}')$

To benchmark the validity of the 2D dipolar quantum simulator, we systematically map out its ground state properties in terms of three experimentally crucial parameters: 1) the depth of the optical lattice potential $A_0$, which we vary from $5 E_R$ to $10 E_R$, 2) the DDI strength $V$, which we vary in the regime of $1 E_R$ to $10 E_R$, and 3) different geometries (varying $S_x$ and $S_y$ independently) and number of particles, which allows us to probe a wide parameter space of filling fractions $\nu$ up to $\nu=0.67$.

%%%%%%%%%%%%%%%%%%%%%%%%%%%%%%%%%%%%%%%%%%%%%%%%%%%%%%%%%%%%%%%%%%%%%%%%%%%%%%%%%%%%%%%%%%%%%%%%%%%%%%%%

%%%%%%%%%%%%%%%%%%%%%%%%%%%%%%%%%%%%%%%%%%%%%%%%%%%%%%%%%%%%%%%%%%%%%%%%%%%%%%%%%%%%%%%%%%%%%%%%%%%%%%%%
\textit{Methods} ---
To model the experimental setup, we solve the many-body Schr\"{o}dinger equation directly for the continuum Hamiltonian~\eqref{eq_continuumH_maintext} by employing the MultiConfigurational Time-Dependent Hartree method for bosons (MCTDH-B)~\cite{Streltsov:2006, Streltsov:2007, Alon:2007, Alon:2008}, implemented by the MCTDH-X software~\cite{Lode:2016,Fasshauer:2016,Lin:2020,Lode:2020,MCTDHX}, which recasts the many-body wave function into a superposition of $M$ single-particle functions called orbitals.
This allows us to obtain the continuum many-body wave function $\ket{\Psi}_{\mathrm{C}}$ and observables derived from it, such as the total energy of the system $E$ and the particle density $\rho_{\mathrm{C}}(\mathbf{x})$~\cite{supmat}.
For a sufficiently deep optical lattice potential, the continuum system should map onto a BHM via a tight-binding approximation where the field operators are rewritten in terms of maximally localized Wannier functions~\cite{Wannier:1937,Marzari:1997,Souza:2001,Modugno:2012,Marzari:2012,Walters:2013,supmat}, $\hat{\Psi}^{(\dagger)}(\textbf{x}) = \sum_{\boldsymbol{\alpha}} w^{(*)}_{\boldsymbol{\alpha}}(\textbf{x}) \hat{b}^{(\dagger)}_{\boldsymbol{\alpha}}$, leading to the lattice Hamiltonian
\begin{align}
\mathcal{H}_{\mathrm{DBH}} &= 
-\sum_{\boldsymbol{\alpha},\boldsymbol{\beta}} J_{\boldsymbol{\alpha}\boldsymbol{\beta}} \hat{b}^{\dagger}_{\boldsymbol{\alpha}} \hat{b}_{\boldsymbol{\beta}} 
+ \sum_{\boldsymbol{\alpha},\boldsymbol{\beta},\boldsymbol{\gamma},\boldsymbol{\delta}} V_{\boldsymbol{\alpha}\boldsymbol{\beta}\boldsymbol{\gamma}\boldsymbol{\delta}} \hat{b}^{\dagger}_{\boldsymbol{\alpha}} \hat{b}^{\dagger}_{\boldsymbol{\beta}} \hat{b}_{\boldsymbol{\gamma}} \hat{b}_{\boldsymbol{\delta}}.
\label{eq:lattice-Ham_maintext} 
\end{align}
Here $\hat{b}^{(\dagger)}_{\boldsymbol{\alpha}}$ denote bosonic operators for band $\sigma$ of site $j$ summarized in a unique index $\boldsymbol{\alpha}=(j,\sigma)$.
The lattice model encapsulates tunneling processes between different sites and intra- or interband density-density interactions, with couplings $J_{\boldsymbol{\alpha}\boldsymbol{\beta}}$ and $V_{\boldsymbol{\alpha}\boldsymbol{\beta}\boldsymbol{\gamma}\boldsymbol{\delta}}$ obtained from Wannier function overlaps 
~\cite{supmat}.

Long-range interactions in closed geometries can quickly lead to higher-band populations.
To discern this, we consider dipolar BHM in the lowest-band approximation (1BDBH) and the ``one-and-a-half band'' approximation (1.5BDBH), where one band is used in one spatial direction and two bands are employed in the other.
To obtain the ground state of the lattice Hamiltonian we perform exact diagonalization with the QuSpin Python library \cite{ED_BH, QuSpin_part_1, QuSpin_part_2}. 
We then use the Wannier function basis to reconstruct continuum versions $\ket{\Psi_{\mathrm{1/1.5BDBH}}}$ with corresponding density $\rho_{\mathrm{1/1.5BDBH}}(\mathbf{x})$.
This procedure in turn allows us to compute many-body fidelities $_{\mathrm{C}}\braket{\Psi}{\Psi}_{\mathrm{1/1.5BDBH}}$ to quantify the agreement between continuum and lattice descriptions~\cite{supmat}.
%%%%%%%%%%%%%%%%%%%%%%%%%%%%%%%%%%%%%%%%%%%%%%%%%%%%%%%%%%%%%%%%%%%%%%%%%%%%%%%%%%%%%%%%%%%%%%%%%%%%%%%%

%%%%%%%%%%%%%%%%%%%%%%%%%%%%%%%%%%%%%%%%%%%%%%%%%%%%%%%%%%%%%%%%%%%%%%%%%%%%%%%%%%%%%%%%%%%%%%%%%%%%%%%%
\textit{Results} ---
We begin by considering systems described by $M=S$ orbitals and Wannier functions (lowest-band), respectively.
Figure~\ref{fig:phase-diagrams} delineates a comprehensive portrayal of fidelity across a spectrum of variables including optical lattice depth, DDI intensity, and filling fractions.
Each graphical representation is derived from a distinct combination of lattice dimensions and configurations ($3 \times 3$, $3 \times 4$, $4 \times 4$, or $5 \times 5$) in conjunction with a range of particle numbers ($N=3$ to $N=6$)~\cite{supmat}. 
Such diversity facilitates the formulation of overarching conclusions concerning the continuum-lattice mapping paradigm. 
Predominantly, Fig.~\ref{fig:phase-diagrams} exhibits a tendency where fidelity diminishes for shallower lattices (lower $A_0$) and intensified DDIs (higher $V$).

The first observation is in accordance with general expectations that the tight-binding mapping is less precise for shallower lattices, since the interband gap is generally proportional to $A_0$.
The second observation unveils a fascinating aspect linked to the extended reach of DDIs.
As $V$ is increased, so does the interaction energy and more and more coupling terms between distant sites have to be accounted for in the lattice picture. 
This encompasses not merely single-band density-density interactions, but also density-assisted tunneling terms across sites and bands, culminating in a larger projection of the continuum ground state onto excited lattice bands.
Consequently, the fidelity of the continuum state with the single-band lattice state is reduced, a phenomenon consistent with prior studies in 1D setups~\cite{Hughes:2023}.

\begin{figure}[!]
\includegraphics[width=0.48\textwidth]{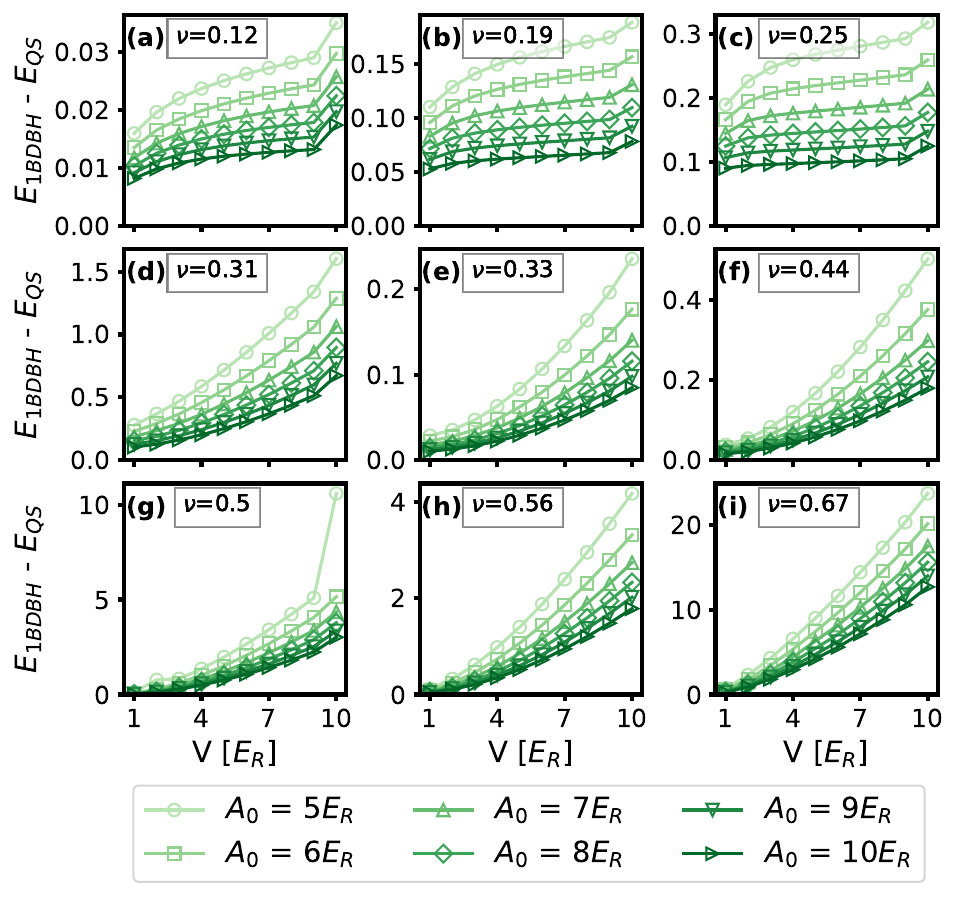}
\caption{\textbf{Energetics of the dipolar quantum simulator.} (a)-(i) Difference in the ground state energy between the 1BDBH model ($E_{1BDBH}$) and the continuum quantum simulator ($E_{QS}$), plotted over increasing filling fractions $\nu$. 
}
\label{fig:energy-diff}
\end{figure}

The increasing discrepancy between continuum and lattice calculations in the low $A_0$ / high $V$ regimes is also corroborated by comparing the energy of the corresponding ground states, shown in Fig.~\ref{fig:energy-diff}.
The continuum ground state energy being systematically lower than the corresponding lattice one reiterates that the continuum description is more fundamental than the mapping to the lattice.
Moreover, as with the fidelity, the difference in energy increases with stronger interactions and shallower lattices, suggesting an increased population of higher bands.

%%%%%%%%%%%%%%%%%%%%%%%%%%%%%
%%%%%%%%%%%%%%%%%%%%%%%%%%%%%

While $A_0$ and $V$ significantly influence the amount of higher-band contribution, what really affects this is the filling fraction $\nu$.
Both Fig.~\ref{fig:phase-diagrams} and Fig.~\ref{fig:energy-diff} show that up to $\nu = 0.25$, the lowest-band dominates the physics with above 99\% fidelity across most scenarios. 
For slightly larger filling fractions up to 0.44, the fidelity remains very high in most regions and only drops to around 90\% at low lattice depths ($A_0 \lessapprox 6 E_r$) and/or very strong interactions ($V \gtrapprox 7 E_R$).
However, filling fractions above 0.5 see a rapid decline in the fidelity to values below 50\%.
Interestingly, this can even occur for deep lattices with $A=10 E_R$.
The reason for this decrease should again be attributed to a sharp increase in the interaction energy -- this time by cramming more particles -- which grants them the ability to overcome the band gap and excite states in higher bands, thereby reducing the lowest-band population.
The dipolar quantum simulator at high densities is thus definitely breaking out of the lowest-band picture.

%%%%%%%%%%%%%%%%%%%%%%%%%%%%%
\begin{figure}[!]
\includegraphics[width=0.48\textwidth]{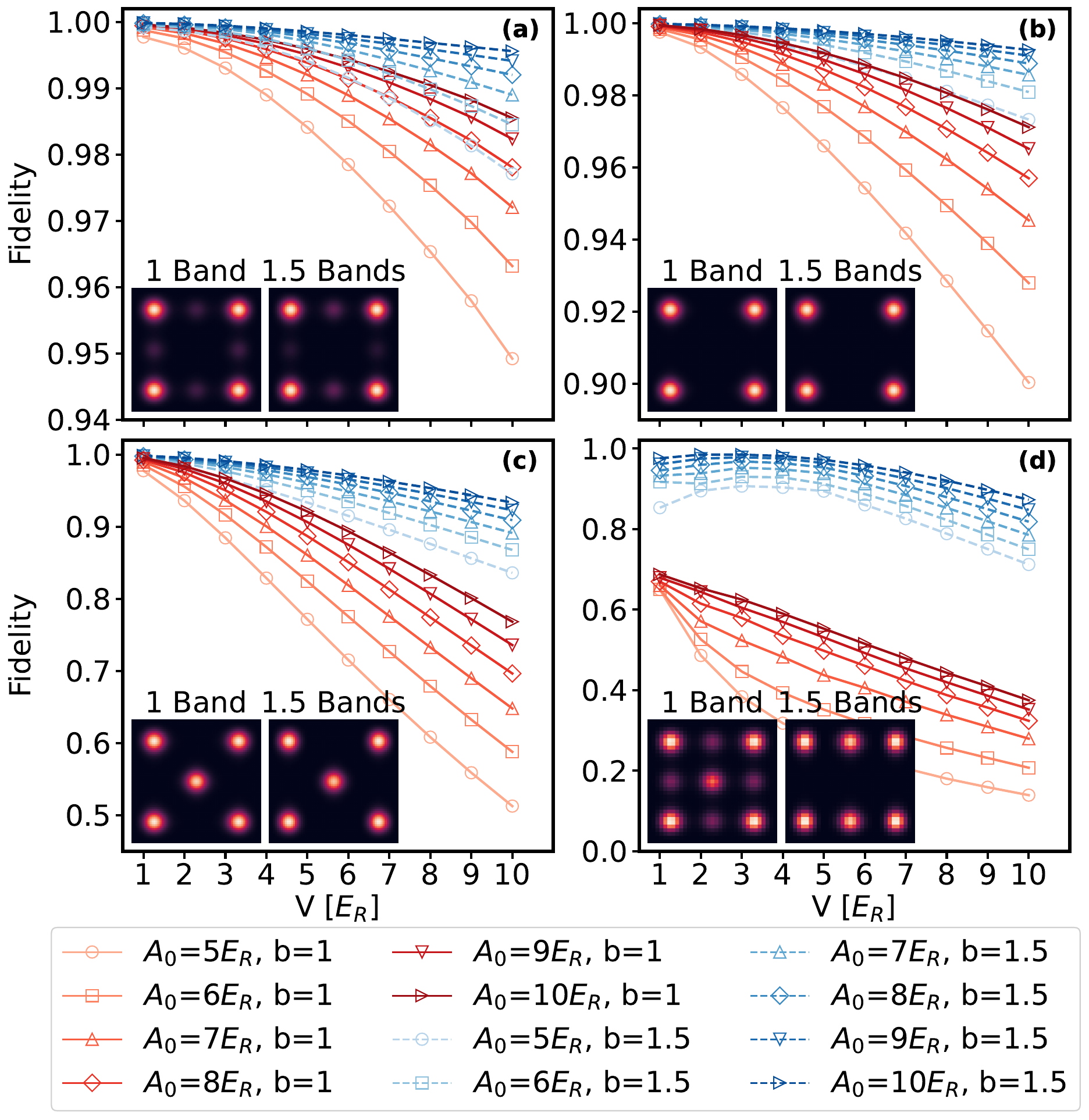}
    \caption{
    \textbf{Improvement of the quantum simulator accuracy for the higher-band model.} Many-body fidelity between continuum and 1-band or 1.5-band lattice model in a $3 \times 3$ geometry as a function of $A_0$ and $V$ for (a) $\nu$=0.33, (b) $\nu$=0.44, (c) $\nu$=0.56, (d) $\nu$=0.67.
    The insets show the particle density of the single- and 1.5-band dipolar BHM for the values of $A_0=5E_R$, $V=10E_R$. 
    }
\label{fig:fidelities}
\end{figure}

%%%%%%%%%%%%%%%%%%%%%%%%%%%%%

To quantify the contributions from higher bands, in Fig.~\ref{fig:fidelities} we present a detailed comparisons between the 1BDBH and the 1.5BDBH models for the $3\times3$ geometry.
In simpler 1BDBH calculations, as previously seen, the fidelity drops monotonically for decreasing $A_0$ and increasing $V$.
Higher filling states with $N=5$ and $N=6$ exhibit the most striking fidelity drop, with low points of 51\% and 14\% respectively.
However, by including additional bands, accuracy improves for all scenarios. 
This improvement is most notable for $N=6$ ($\nu = 0.67$), where the two models predict completely different states as can be seen in the insets of Fig.~\ref{fig:fidelities}(d): the ground state of the 1BDBH model exhibits a checkerboard pattern, while the one of the 1.5BDBH is instead in a striped configuration. 
This illustrates two key points:
Approximating the dipolar quantum simulator with lowest-band models can lead to significant errors in high-density situations.
Conversely, the setup is already equipped to handle richer, higher band models without extra adjustments.

\begin{figure}[!]
\includegraphics[width=0.48\textwidth]{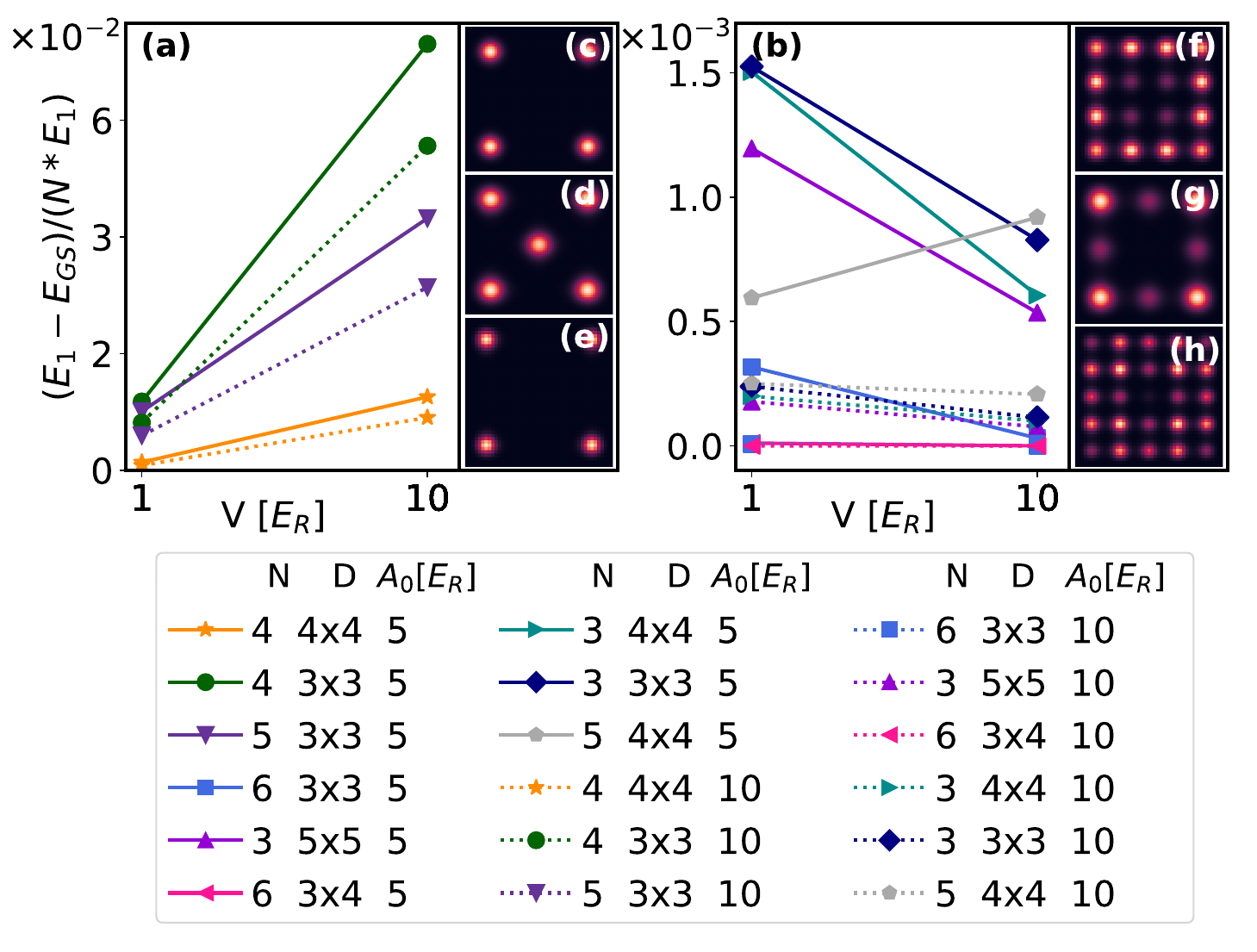}
    \caption{\textbf{Effect of lattice depth and dipolar interaction strength on incommensurability.} Relative energy differences per particle ($N$) between the $1^{\text{st}}$ excited state ($E_1$) and the ground state ($E_{GS}$), quantifying incommensurability of the given geometry and filling fraction ($\nu$). 
    The insets show the 1BDBH model's particle density of the commensurable states for (c) $\nu$=0.44, $A_0$=10$E_R$, V=10$E_R$, (d) $\nu$=0.56, $A_0$=5$E_R$, V=1$E_R$, (e)  $\nu$=0.25, $A_0$=5$E_R$, V=10$E_R$, and of the incommensurable states for (f) $\nu$=0.19, $A_0$=5$E_R$, V=1$E_R$, (g) $\nu$=0.33, $A_0$=5$E_R$, V=1$E_R$, (h) $\nu$=0.12, $A_0$=5$E_R$, V=1$E_R$.}
\label{fig:lattice-energies}
\end{figure}

Another key factor that affects the properties of the ground state in the quantum simulator is system geometry.
Depending on particle number and lattice geometry, strong repulsion may make the particle ordering incommensurate.
This can lead to spontaneous lattice symmetry breaking which generates quasidegenerate ground states.
A summary of the lattice quasidegeneracies is shown in Fig.~\ref{fig:lattice-energies}, quantified as the gap between the ground state and the first excited state for all the geometries considered at the four extremal values in our $(V,A_0)$-parameter range. 
Two clusters of states can be identified, with the clustering itself being essentially independent of the values of $V$ and $A_0$, thereby reflecting the underlying geometry.
Within each cluster, $V$ has opposite effects (stabilizing vs. destabilizing).
The first cluster consists of $N=4,5$ in the $3\times3$ and $N=4$ in the $4\times4$ lattice, i.e. geometries that can accommodate the ground state in the most symmetric configuration in the localized limit (four isolated particles in the corners and none or one in the center, respectively).
These states preserve all discrete symmetries of the lattices, such as $C_4$ rotations, reflections, and inversions, which makes them very stable as confirmed by a large gap to excited states.  
The second cluster, instead, encompasses geometries with incommensurability.
Examples include all the $N=3$ states and the $N=5$ state in the $4 \times 4$ lattice. 
For these states, localizing forces such as lattice potential and repulsions clash with the available minima provided by the lattice symmetries.
In milder cases (e.g. $N=3$, $5 \times 5$ lattice), this favors a more delocalized ground state which does accommodate the symmetries but can be easily excited to similar states of quasidegenerate energies~\footnote{The half-filled $S_x \times S_y = 3 \times 4$ case despite being commensurate was added to the incommensurate cluster due to its completely degenerate ground state.}.
In more severe cases (e.g. $N=6$, $3 \times 3$ and $3 \times 4$ lattices), a total breaking of rotational symmetries occurs, leading to a ground state partitioned into multiply-degenerate sectors, e.g. two opposite checkerboard patterns, a feature known in extended BH models for states above half filling~\cite{Trefzger:2011,Maik:2013,Suthar:2020}.
These results demonstrate that DDI not only offer an experimental platform to realize multiband models, but in conjunction with different geometries can also produce ground-state (quasi)degeneracies that are important ingredients to achieve exotic quantum states i.e. with topological features.

%%%%%%%%%%%%%%%%%%%%%%%%%%%%%
%%%%%%%%%%%%%%%%%%%%%%%%%%%%%

%%%%%%%%%%%%%%%%%%%%%%%%%%%%%%%%%%%%%%%%%%%%%%%%%%%%%%%%%%%%%%%%%%%%%%%%%%%%%%%%%%%%%%%%%%%%%%%%%%%%%%%%

%%%%%%%%%%%%%%%%%%%%%%%%%%%%%%%%%%%%%%%%%%%%%%%%%%%%%%%%%%%%%%%%%%%%%%%%%%%%%%%%%%%%%%%%%%%%%%%%%%%%%%%%
\textit{Discussion and Outlook} ---
Dipolar molecules and magnetic atoms are becoming a staple in ultracold atomic labs around the world.
By transferring population to the ground state using STIRAP and controlling their large dipole moment via electric field alignment, their interaction regimes can reach $\approx 20$kHz at a $532$nm separation, i.e. $15 E_R$ for a lattice wavelength $\lambda = 1064$nm~\cite{Stevenson:2023}. 
Optical lattices with hard-wall boundaries can be realized by adding flat-bottom traps to counter-propagating laser beams~\cite{Gaunt:2013,Mazurenko:2017,Gall:2021,Navon:2021}, and lattice depths of up to $10 E_R$ are standard.
Thus, all the systems we have studied should be experimentally accessible in near-term experiments. 

We have presented a comprehensive analysis of the accuracy of 2D dipolar quantum simulators by performing quantitative comparisons between continuum models of experimental setups and effective lattice models in a wide range of relevant lattice depths, interaction strengths, and filling fractions.
Our study highlights that ultracold dipolar systems can be a very powerful instrument to reach regimes of multiband occupation and thereby allow to probe more realistic, complex long-range interacting lattice models.
The population of higher bands can drastically alter ground-state properties.
This is exemplified for $N=6$ particles in a $3 \times 3$ geometry, where the 1.5BDBH model predicts a striped state (with anisotropic spatial order), but the 1BDBH model gives rise to a checkerboard pattern.
In general, regimes of high interaction strengths, shallow lattices, and high fillings should generate the richest multiband physics.
While highly symmetric states could be simulated with greater accuracy, incommensurability effects emerging from the competition between spatial geometry and localizing interactions and potentials can drastically alter the spectral landscape.
As a consequence, quasidegenerate ground states can arise, which can be completely modified by the presence of higher-band contributions.
These findings should even prompt the question whether lowest-band descriptions are sufficient to encompass the most exotic physics of lattice models systems, for instance in solid state settings, which interact via even longer ranged couplings such as unscreened or weakly screened Coulomb forces, e.g. in low-dimensional or thin-film systems. 
Our results showcase the enormous possibilities offered by ultracold dipolar quantum simulators to reproduce these multiband lattice models.

Our study, based on a rigorous, quantitative comparison between lattice and continuum descriptions, should provide a reliable blueprint for the correct realization of long-ranged quantum simulators.
While we have focused on repulsive interactions in two dimensions, we expect a rich physics to arise for attractive regimes and different spatial geometries, too.
This is particularly important in light of discoveries of interesting new phases in attractive dipolar BHM models such as exotic quantum liquids~\cite{Morera:2023,Marciniak:2023,Staudinger:2023} and self-bound quasicrystalline order~\cite{Kora:2020}.
In particular, a key question to address will be the treatment of singularities arising for DDIs at zero distance, which can occur in the continuum but are truncated by the lattice formulation.
Our approach of quantitatively comparing continuum and lattice descriptions by calculating many-body overlaps should shed light on those issues.

\acknowledgements
We acknowledge computation time on the ETH Zurich Euler cluster and at the High-Performance Computing Center Stuttgart (HLRS).
We thank O. Alon, E. Bergholtz, R. Chitra, D. Jaksch, and P. Nayak for useful comments on the manuscript.
This work is supported by the Swedish Research Council (2018-00313) and Knut and Alice Wallenberg Foundation (KAW) via the project Dynamic Quantum Matter (2019.0068) as well as U.K. Engineering and Physical Sciences Research Council (EPSRC) Grants no. EP/P01058X/1 (QSUM) and no. EP/P009565/1 (DesOEQ).

\bibliography{lattice-MCTDH-X_bib}

\end{document}